# Graphene/α-RuCl₃: An Emergent 2D Plasmonic Interface

Daniel J. Rizzo[1], Bjarke S. Jessen[1,2], Zhiyuan Sun[1], Francesco L. Ruta[1,3], Jin Zhang[4], Jia-Qiang Yan[5,6], Lede Xian[4], Alexander S. McLeod[1], Michael E. Berkowitz[1], Kenji Watanabe[7], Takashi Taniguchi[8], Stephen E. Nagler[9], David G. Mandrus[5,6], Angel Rubio[4,10,11], Michael M. Fogler[12], Andrew J. Millis[1,10], James C. Hone[2], Cory R. Dean[1], D.N. Basov[1],*

[1]Department of Physics, Columbia University, New York, NY, 10027, USA

[2]Department of Mechanical Engineering, Columbia University, New York, NY, 10027, USA

[3]Department of Applied Physics and Applied Mathematics, Columbia University, New York, NY, 10027, USA

[4]Department of Physics, Max Planck Institute for Structure and Dynamics of Matter and Center for Free-Electron Laser Science, 22761 Hamburg, Germany

[5]Materials Science and Technology Division, Oak Ridge National Laboratory, Oak Ridge, Tennessee 37831, USA

[6]Department of Materials Science and Engineering, University of Tennessee, Knoxville, Tennessee 37996, USA

[7]Research Center for Functional Materials, National Institute for Materials Science, 1-1 Namiki, Tsukuba 305-0044, Japan

[8]International Center for Materials Nanoarchitectonics, National Institute for Materials Science, 1-1 Namiki, Tsukuba 305-004, Japan

[9]Neutron Scattering Division, Oak Ridge National Laboratory, Oak Ridge, Tennessee 37831, USA

[10]Center for Computational Quantum Physics, Flatiron Institute, New York, New York 10010, USA

[11]Nano-Bio Spectroscopy Group, Universidad del País Vasco, San Sebastian 20018, Spain

[12]Department of Physics, University of California San Diego, La Jolla, CA, 92093, USA

*Correspondence to: db3056@columbia.edu

**Work function-mediated charge transfer in graphene/α-RuCl₃ heterostructures has been proposed as a strategy for generating highly-doped 2D interfaces. In this geometry, graphene should become sufficiently doped to host surface and edge plasmon-polaritons (SPPs and EPPs, respectively). Characterization of the SPP and EPP behavior as a function of frequency and temperature can be used to simultaneously probe the magnitude of interlayer charge transfer while extracting the**

**optical response of the interfacial doped α-RuCl₃. We accomplish this using scanning near-field optical microscopy (SNOM) in conjunction with first-principles DFT calculations. This reveals massive interlayer charge transfer ($2.7 \times 10^{13}$ cm$^{-2}$) and enhanced optical conductivity in α-RuCl₃ as a result of significant electron doping. Our results provide a general strategy for generating highly-doped plasmonic interfaces in the 2D limit in a scanning probe-accessible geometry without need of an electrostatic gate.**

Research in two-dimensional (2D) materials has generated a pantheon of tailor-made atomically-thin layers whose optoelectronic properties are sensitive to their lattice symmetry and elemental composition. The behavior of low dimensional materials can be further tuned via chemical or electrostatic doping (*1-4*). Moreover, van der Waals (vdW) heterostructures composed of different 2D layers can possess emergent behavior based on interfacial effects, including moiré-dependent orbital hybridization (*5-11*), magnetic interactions (*12*), and proximal superconductivity (*13-15*). Recent theoretical (*16, 17*) and experimental (*18-20*) studies have proposed work function-mediated interlayer charge-transfer as yet another route to realizing novel interfacial 2D behavior. Specifically, the relatively large work function of α-RuCl₃ ($\Phi_{RuCl3}$ = 6.1 eV) (*18*) compared to graphene ($\Phi_g$ = 4.6 eV) (*21*) leads to interlayer charge transfer that mutually dopes each material in excess of $n = 10^{13}$ cm$^{-2}$ holes and electrons in graphene and α-RuCl₃, respectively. Such large charge carrier densities will likely impact the magnetic correlated insulating ground state of α-RuCl₃ (*22-27*), and may influence the presumed quantum spin liquid (QSL) it hosts in the 2D limit (*28-30*). On the other hand, this interlayer charge transfer should sufficiently dope graphene to provide the necessary conditions for generating interfacial plasmon polaritons without the need of electrical contacts or chemical alteration (as required previously). Plasmon polaritons are hybrid light matter modes formed by infrared photons and Dirac electrons that can travel over macroscopic distances in structures with high mobility (*31, 32*). The temperature and frequency dependent properties of these confined plasmons depend sensitively on the local dielectric and charge environment. Therefore, a plasmonic probe of graphene/α-RuCl$_3$ heterostructures would permit a quantitative analysis of the interlayer charge dynamics with nanoscale spatial resolution.

Here we harness surface plasmon polaritons (SPPs) in graphene to investigate the charge-transfer interface formed by graphene/α-RuCl$_3$. The wavelength of SPPs is dictated by the electron density in the host medium (*33*) thereby allowing us to read the strength of the charge transfer in our graphene/α-RuCl$_3$ heterostructure directly from the images of polaritonic fringe patterns. At the same time, sharp boundaries in the conductivity along the edge of doped graphene host 1D edge plasmon polaritons (EPPs) that disperse based on the spatial profile of the graphene/α-RuCl$_3$ charge transfer near the graphene edges (*34-36*). Therefore, the collective behavior of SPPs and EPPs in graphene/α-RuCl$_3$ heterostructures permits us to characterize the magnitude and spatial dependence of the interlayer charge transfer. Moreover, the electric field of the polaritonic wave extends over tens of nanometers outside of the graphene plane, and offers unrestricted access to the charge dynamics of electrons transferred into α-RuCl$_3$.

A polaritonic probe of graphene/α-RuCl$_3$ heterostructures is enabled by scattering-type scanning near-field optical microscopy (s-SNOM), which permits us to map the near-field optical response of this interface at sub-diffractional length scales. Thus, s-SNOM is well suited to image SPP waves over a wide range of IR frequencies (here, $\omega = 900 - 2300$ cm$^{-1}$). SPP imaging has allowed us to reconstruct the energy momentum dispersion $\omega(q)$ associated with an exceptionally high charge-carrier density in graphene ($n = 2.7 \times 10^{13}$ cm$^{-2}$, $|\mu| = 0.61$ eV) that we assign to work function-mediated charge transfer with α-RuCl$_3$. In addition, we commonly observe point-like topographic defects in the interior of graphene microcrystals where charge transfer is suppressed, triggering characteristic concentric SPP fringes. Analysis of the EPP dispersion provides further insight into the spatial profile of interlayer charge transfer near the graphene edge.

We are also able to extract the response function of highly doped α-RuCl$_3$ by modelling the real and imaginary part of the complex SPP wavevector $q = q_1 + iq_2$ that we investigated as a function of both frequency and temperature. Here, $q_1$ encodes information about the plasmon wavelength, while $q_2$ characterizes the plasmon damping length scales. If the dielectric environment of graphene is known, the quality factor $Q = q_1/q_2$ can be used to extract the scattering rate of electrons in graphene. After accounting for known sources of damping, we observe residual SPP scattering that is consistent with a non-trivial optical response in the topmost layer of α-RuCl$_3$ at all experimental frequencies. This conclusion is well supported by first-principles density functional theory (DFT) calculations of model graphene/α-RuCl$_3$ heterostructures, which shows charge being largely confined to the topmost layer of α-RuCl$_3$ occupying t$_{2g}$ bands.

**Observation of Plasmon-Polaritons in hBN/Graphene/α-RuCl$_3$ Heterostructures**

In order to explore the charge dynamics of the graphene–α-RuCl$_3$ interface, we first constructed a hexagonal boron nitride (hBN)-encapsulated graphene/α-RuCl$_3$ heterostructure on a SiO$_2$/Si substrate as shown in Fig. 1A (see supplementary materials and Fig. S1 for assembly details) (*37*). We include a few layers of encapsulating hBN to protect the graphene surface from unwanted doping via atmospheric contaminants (*38*) while ensuring that it is sufficiently thin to permit efficient coupling of our optical scanning probe with the underlying graphene.

s-SNOM images performed on graphene/α-RuCl$_3$ heterostructures reveal three distinct but related forms of polaritonic waves (Fig. 1B). Here, oscillations in the near-field amplitude are observed emanating inward from the graphene edge that are characteristic of SPPs. We also observe SPP fringes emanating radially from point-like ($r <$ 50 nm, $h \sim$ 1 nm) topographic defects that appear to be acting as local plasmon reflectors. Finally, we observed EPP fringes oscillating along the graphene edge. Analyses of the dispersions (i.e. $q_1$ vs. $\omega$) of all three plasmonic features each reveal a unique facet of the graphene/α-RuCl₃ charge transfer process. Furthermore, analysis of the plasmonic damping (i.e. $q_2$) permits us to interrogate the effect of this charge-transfer on the optical response of the underlying doped α-RuCl₃.

**Plasmon-Polaritons as Probe of Interlayer Charge Transfer**

In order to accurately account for the depth of interlayer charge transfer in our modelling of the graphene/α-RuCl₃ optical response, we perform DFT calculations on two heterostructures containing one (1L) and two (2L) layers of α-RuCl₃ (Figs. 1C, D). Our modelling of the α-RuCl₃ band structure accounts for different values of the Hubbard U for the Ru 4d and Cl 3p orbitals ($U_{4d}$ = 1.96 eV and $U_{3p}$ = 5.31 eV, respectively) (Fig. S2) (*37*). For the 1L and 2L heterostructures, a large positive shift in the graphene Fermi energy, $E_F$, is observed (0.54 eV and 0.61 eV, respectively) relative to that of pristine graphene. This shows that the addition of an α-RuCl₃ layer does not substantively impact the expected doping level of the adjacent graphene (Figs. 1C and D). Analysis of the Bader charge (*39*) of the 2L heterostructures shows that the interfacial layer of α-RuCl₃ is nearly as heavily doped as it is in the 1L structure (0.069 versus 0.063 $|e|$/Ru atom, respectively), while the second layer of α-RuCl₃ receives <10% of the charge carriers of the first (Figs. 1C, D). DFT-calculations further suggest that the carrier density of α-RuCl₃ can be effectively modulated by including *h*BN layers between graphene and α-RuCl3 as exhibited in Fig. S2 (*37*). Together, these observations reveal that the interfacial layer of α-RuCl₃ is sufficiently doped to experience significant changes in frontier electronic states, while subsequent layers more closely resemble the undoped electronic structure. With this in mind, all subsequent models of the optical response of our stack presented in this text assume that the top-most layer of α-RuCl$_3$ is highly doped, while all subsequent layers are treated as undoped.

We begin our experimental analysis by imaging edge-launched SPP fringes over a wide range of frequencies in the mid-IR using s-SNOM (Figs. 2A, B). No such fringes were observed for graphene residing directly on SiO$_2$ (Fig. S3) (*37*), suggesting that such behavior originates from the graphene/α-RuCl$_3$ interface. The SPP wavelength is dependent on the incident laser frequency, $\omega$, being tuned by more than an order of magnitude over the experimental frequency window (Fig. 2B). By fitting the average line profile of the SPP fringes with a combination of plane- and cylindrical-wave terms (Fig. S4) (*32, 37, 40*), we are able to obtain the experimental dispersion for SPPs generated at the graphene/α-RuCl$_3$ interface (Fig. 2C). The experimental dispersion shows two branches: a lower branch located over the frequency range 900–1010 cm$^{-1}$ and an upper branch spanning 1380–2300 cm$^{-1}$ separated by a region of SiO$_2$ and hBN phonons, in accord with prior results (*41*).

An evident novelty of data in Fig. 2 is that the plasmonic dispersion is achieved without resorting to either gating or chemical doping. The appearance of SPP fringes in graphene/α-RuCl$_3$ therefore attests to substantial charge transfer between graphene and α-RuCl$_3$. We are able to further quantify the magnitude of the charge transfer by calculating the theoretical plasmon dispersion from the p-polarized reflection coefficient ($r_p$) and comparing it to the experimental plasmon dispersion. Our model properly considers the response functions and thicknesses of all constituent layers in encapsulated graphene/α-RuCl$_3$ heterostructure as described in the supplementary discussion (*37*). Calculations of the loss function, Im($R_p$), validate that the dominant contribution originates from plasmons in the graphene layer. A least-squares regression between the experimental dispersion and maxima in the loss function yield the graphene chemical potential $|\mu|$ as a fitting parameter (Figs. 2C, S5) (*37*). The model captures all the trends in the data with the best-fit value of $|\mu| = 0.61$ eV corresponding to $n = 2.7 \times 10^{13}$ cm$^{-2}$ carriers in the graphene layer due to interlayer charge transfer (Fig. 2C). This value is consistent with previous measurements of the Hall resistance ($n = 2.8 \times 10^{13}$ cm$^{-2}$)) (*18*), and is in good agreement with theoretical expectation (Figs. 1C, D) (*16*). We note that the agreement between the experimental and model dispersions was greatly improved by assuming that the top-most layer of α-RuCl$_3$ possesses additional optical conductivity compared to that of undoped α-RuCl$_3$ (Fig. S5) (*42, 43*). Hence, our data support the notion of large interfacial doping and enhanced optical conductivity in α-RuCl$_3$ as suggested by DFT calculations (Figs. 1C, D) (*37*).

While graphene/α-RuCl$_3$ SPP fringes were primarily observed launching and reflecting from the graphene edge, the interior of graphene reveals a different type of plasmonic oscillations exactly centered around topographic point defects (Fig. 3A). We posit that SPPs reflect off these defects, forming radially-symmetric patterns in the near-field amplitude (Figs. 3A, B, S6) (*37*). Detailed modeling corroborates this intuition. Specifically, we use approximate solutions to the Helmholtz equation screened by long-range Coulomb interactions to model the experimental near-field amplitude profile as a function of the radial distance from the defect (see supplementary materials for more details) (*37*). Here, the defect is described as a circular region with a variable conductivity relative to the uniform areas in the interior of the sample (Figs. 3C, D), while the near-field amplitude is approximated by a position-dependent reflectivity for illuminating fields from a raster-scanned point dipole. We observe that the sign and magnitude of the primary SPP fringe depends on the defect conductivity relative to that of the surrounding graphene.

Conductivity-excess defects show a bright primary fringe, while conductivity-depletion defects show a dark primary fringe. As is evident from Fig. 3C, a good fit to the experimental fringe cross-section can only be achieved with a maximally depleted (i.e. non-conducting) region at the defect site. Indeed, all experimentally observed defects at all frequencies possess a dark primary fringe, and are therefore all considered to be conductivity-depletion defects. Agreement between plasmon modelling and experiment is optimized with a defect profile following a Lorentzian distribution with a radius of 40 nm, suggesting that charge depletion only occurs roughly within the bounds of the topographic defects (Fig. 3B). Using this model for the near-field amplitude in the vicinity of a charge depletion defect, we are able to extract the experimental defect-SPP dispersion from near-field images of a point defect taken over the frequency range $\omega = 900 - 1720$ cm$^{-1}$ (Figs. 2C, S6) (*37*). The defect-SPPs follow a dispersion that is quite similar to that of edge-launched SPPs for $\omega$ < 1550 cm$^{-1}$, implying a similarly high level of hole doping in the vicinity of the topographic defects. Therefore, the near-field behavior of topographic defects reveals that the defects themselves are undoped, while the graphene in the immediate vicinity is heavily doped. This is consistent with the picture that the topographic defects are small bubbles in the graphene where the interface with α-RuCl$_3$ is not sufficiently well-formed to allow interlayer charge transfer to take place.

The magnitude and spatial-dependence of the interlayer charge transfer between graphene and α-RuCl$_3$ revealed by s-SNOM experiments is further corroborated by Raman spectroscopy (Fig. S7) (*37*). Here, analysis of the graphene G and 2D peak shifts shows the coexistence of a highly-doped ($n$ = 2.5×10$^{13}$ cm$^{-2}$), uniformly strained ($\varepsilon$ = –0.2 %) phase, and an undoped, randomly strained phase. The latter undoped phase only appears to be represented in regions observed to have a high density of point defects in near-field imaging (Fig. S7) (*37*).

While SPP dispersions act as sensitive probes of the interlayer charge transfer in the graphene bulk, the EPP dispersion can provide information about the charge transfer profile near the graphene edge (*34-36*). In order to exploit this, we extract the profiles of the near-field oscillations along the graphene edge as a function of $\omega$ (Figs 4A, B). As with the SPP fringes, the wavelength of the EPP fringes has a clear dependence on $\omega$ as shown in the EPP dispersion (Fig. 4C). As expected (*34-36*), EPPs possess a systematically higher value of $q_1$ compared to SPPs (Fig. 4C). Following the procedure in (*36*), we model the EPP dispersion based on the assumption of a discontinuous jump in the graphene conductivity coinciding with the graphene edge (Fig. 4C). We find that the experimental EPP dispersion is in excellent agreement with this model, suggesting that the substantial charge transfer between graphene and $\alpha$-RuCl$_3$ occurs uniformly up to and including the graphene edge, forming sharp boundaries in the graphene conductivity relative to the SPP wavelength.

**Probing the Optical Properties of Highly-doped $\alpha$-RuCl$_3$ via SPP Dynamics**

We now seek to characterize the optical response of the graphene/$\alpha$-RuCl$_3$ interface by comparing the experimentally-derived plasmon scattering to all known loss channels. The inverse of the SPP quality factor, $Q^{-1}$, can be expressed in terms of the effective dielectric environment, and the optical conductivity of the graphene/$\alpha$-RuCl$_3$ interface:

$$Q^{-1} = \frac{q_2}{q_1} \approx \frac{\sigma_1}{\sigma_2} + \frac{\epsilon_2}{\epsilon_1} = \frac{\gamma(\omega)}{\omega} + \frac{\epsilon_2}{\epsilon_1} \qquad (1)$$

where $\sigma = \sigma_1 + i\sigma_2$ is the complex optical conductivity of the graphene/$\alpha$-RuCl$_3$ interface and $\epsilon = \epsilon_1 + i\epsilon_2$ is the effective dielectric of the environment. The frequency-dependent scattering rate, $\gamma(\omega)$, has several additive components, which can be broadly attributed to either the graphene or $\alpha$-RuCl$_3$ side of the interface:

$$\gamma(\omega) = \gamma_{RuCl_3}(\omega) + \gamma_g(\omega) \qquad (2)$$

Dissipation channels in graphene are well understood, and include two acoustic and three optical graphene phonons (*31*). In principle, additional scattering channels may emerge as a result of non-trivial optical conductivity in the interfacial layer of $\alpha$-RuCl$_3$, which in turn should lead to a suppression of the experimental $Q$. With this in mind, we first extract the plasmon quality factor as a function of $\omega$ from the experimental fits to the near-field SPP fringes shown in Fig. 2B, revealing a systematic decrease in $Q$ with $\omega$ (Fig. 5A). Using (1) along with known substrate optical parameters, the experimental $Q$ can be used to obtain an experimental scattering rate (Fig. 5B) (*37*). The collective contribution of all graphene phonons to the scattering rate is shown in Fig. 5B. The sum of all intrinsic sources of scattering in graphene clearly falls short of the corresponding experimental values, revealing the presence of additional sources of scattering at the graphene/$\alpha$-RuCl$_3$ interface. While disorder is a potential source of additional scattering, such channels tend to contribute frequency- and temperature-independent losses. The clear frequency dependence of the residual experimental scattering suggests that it is not due to simple disorder, and should be attributed to emergent optical conductivity in the doped interfacial $\alpha$-RuCl$_3$. Analysis of this residual scattering (*37*) permits us to extract Re($\sigma$) for the topmost layer of $\alpha$-RuCl$_3$, yielding typical values between 0.02–0.1 $e^2/\hbar$ (Fig. 5C). We assign a model optical response to the topmost layer of $\alpha$-RuCl$_3$ that incorporates both temperature- and frequency- dependent scattering. The addition of this model $\alpha$-RuCl$_3$ loss channel improves the agreement between the theoretical and experimental scattering rates and quality factors. The excess scattering (i.e. $\alpha$-

RuCl$_3$ conductivity) at high frequency suggests that previously observed optical transitions near 0.2 eV persist in α-RuCl$_3$ upon doping (*26*). This analysis of the SPP damping provides evidence that the α-RuCl$_3$ possesses free charge carriers upon doping from interlayer charge transfer.

To further test the possibility of an emergent optical response in doped α-RuCl$_3$, we obtained the temperature dependence ($T = 60 - 300$K) of the SPP damping rate for $\omega = 898$ cm$^{-1}$ (Fig. 5D, E). We observe that the SPP $Q$ increases by more than a factor of three as the sample temperature is brought from 300 K to 60 K (Fig. 5F). Once again, the calculated experimental scattering rate is severely underestimated by phonon scattering and dielectric losses alone, and is only reconciled with theory once the topmost layer of α-RuCl$_3$ is endowed with a non-zero optical response (Figs. 5F, S8). The enhanced scattering rate at all observed temperatures further suggests the presence of active carriers in α-RuCl$_3$.

**Discussion**

Our joint experimental and theoretical plasmonic imaging study of graphene/α-RuCl$_3$ heterostructures provides conclusive evidence of massive work function-mediated charge transfer between these two materials. Moreover, our scanning probe approach reveals the real-space variations of this charge transfer process and allows us to characterize the conductivity of graphene nanobubbles speculated to exist in previous works (*18*). Concurrent analysis of the temperature- and $\omega$-dependence of SPP losses provides a complimentary probe of the doped α-RuCl$_3$ optical properties, and supports the prediction (*16*) that α-RuCl$_3$ possesses emergent optical conductivity due to interlayer charge transfer.

The experimental techniques presented in this work have broad implications in the study of α-RuCl$_3$, graphene plasmonics, and 2D materials generally. The ability to readily detect charge-depletion regions in graphene/α-RuCl$_3$ using s-SNOM provides a robust tool for identifying well-coupled regions of charge-transfer interfaces with graphene. At the same time, the recognition that massive charge transfer is taking place over large uninterrupted areas ( > 10 μm$^2$) of the graphene provides unambiguous validation of the notion that work function-mediated charge transfer can be used to heavily dope 2D materials into new physical regimes. In the past, such high charge carrier densities in graphene could only be achieved with ion liquid gel devices (*44*), or an additional top-gate layer (*45*), which forbid detailed surface studies with a scanning probe. Therefore, our experimental system provides a roadmap for generating record high charge carrier densities in 2D materials while maintaining an open surface for interrogating real-space features at the nm and sub-nm scale with a probe tip. Furthermore, the observation that the graphene optical conductivity forms sharp boundaries wherever the graphene/α-RuCl$_3$ interface is interrupted suggests that α-RuCl$_3$ can be used to create nanoscale graphene p-n junctions without the need of a split local back gate (*46*). Similarly, patterned α-RuCl$_3$ arrays should mimic the sharp periodic potentials used previously to realize photonic crystals in gated graphene/semiconductor architectures (*47*), but without the need of an external power source. Finally, we have introduced a standard procedure for probing the optical response of 2D material interfaces via the SPP damping, providing a route to describe nascent 2D materials that do not possess the large-scale uniformity necessary to be characterized in the far-field.

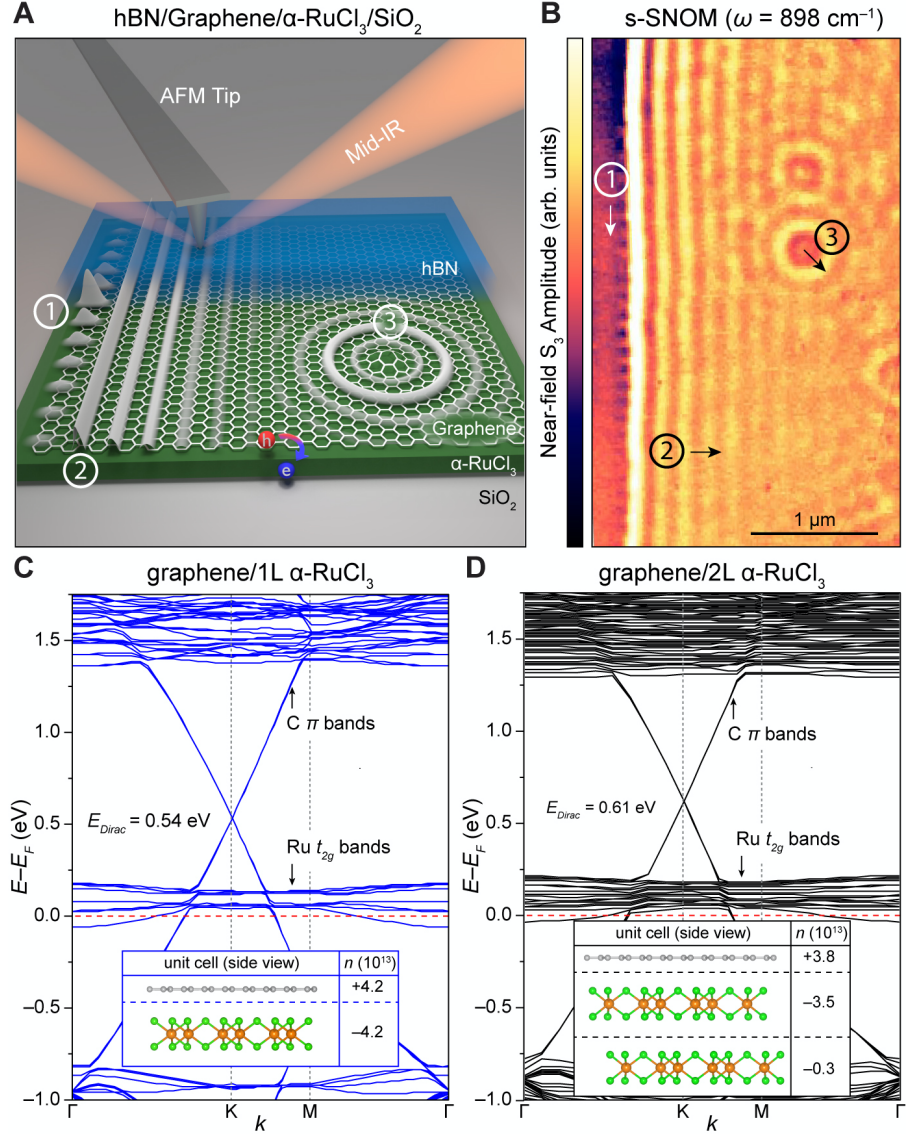

**Fig. 1. Characterization of interlayer charge transfer in graphene/α-RuCl₃ heterostructures using s-SNOM and DFT calculations.** (**A**) Diagram of s-SNOM measurements being performed on a hBN/Graphene/α-RuCl₃/SiO₂ stack. The large difference in the work functions of graphene and α-RuCl₃ leads to substantial hole-doping of graphene, thus providing the necessary conditions for generating and imaging three types of plasmon features: (1) EPPs, (2) edge-launched SPPs, and (3) defect-reflected SPPs. (**B**) Experimental map of the near-field amplitude near the edge of graphene in a hBN/Graphene/α-RuCl₃/SiO₂ heterostructure ($\omega$ = 898 cm⁻¹, $T$ = 60 K) showing

oscillations that are characteristic of the features shown in (A). (**C**) DFT+U+SOC band structure for graphene/1L α-RuCl$_3$ model heterostructure (unit cell shown in the inset). Bands derived from carbon $\pi$ orbitals and Ru $t_{2g}$ orbital are indicated. Inset: The calculated Bader charge in each layer of the model unit cell is indicated in terms of the resulting charge carrier concentration, *n* in units of cm$^{-2}$. (**D**) Same as (C), but for the model heterostructure with two layers of α-RuCl$_3$ (graphene/2L α-RuCl$_3$). Here, the interfacial layer of α-RuCl$_3$ possesses 91% of the electrons transferred from graphene, while the second layer of α-RuCl$_3$ contains only 9%.

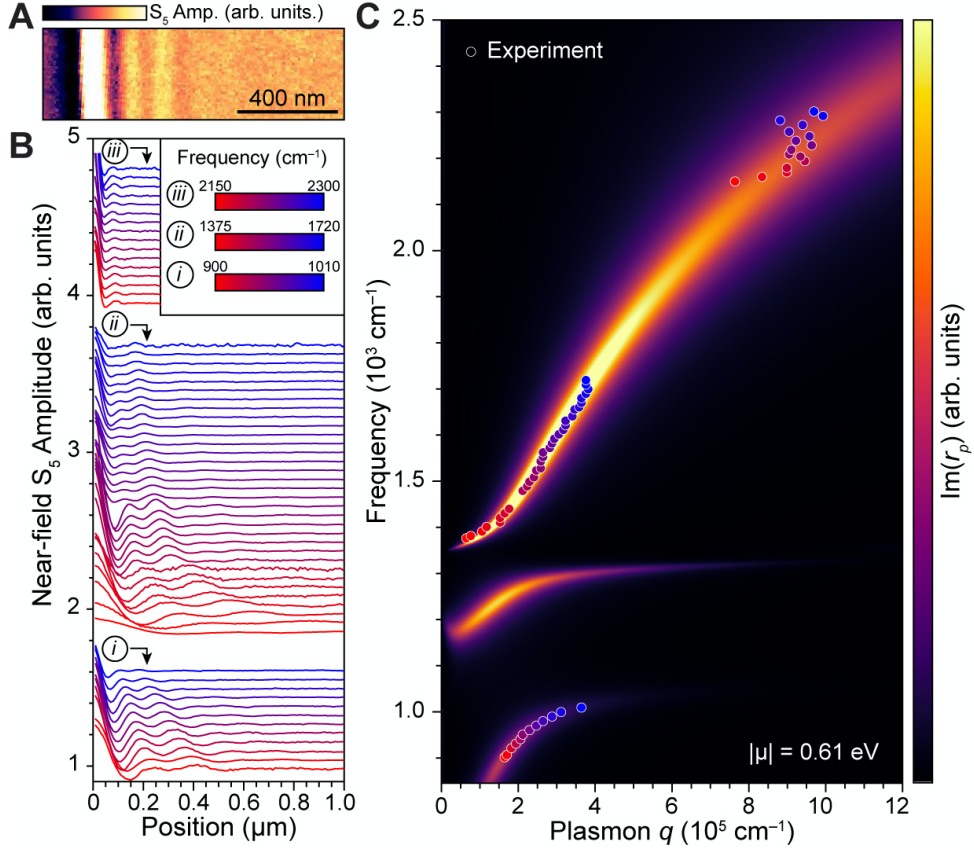

**Fig. 2. Characterization of the SPP dispersion in graphene/α-RuCl₃ heterostructures using s-SNOM.** (**A**) Map of the near-field amplitude ($\omega$ = 960 cm$^{-1}$) near a graphene edge in graphene/α-RuCl₃ heterostructures, showing oscillatory behavior characteristic of graphene SPP fringes. (**B**) Line profiles of the average near-field amplitude as a function of distance from the graphene edge for a broad range of frequencies ($\omega$ = 900 – 2300 cm$^{-1}$), showing substantial shifts in the SPP wavelength. Here, sequential curves are offset vertically by 0.1 for clarity and grouped based on the three different ranges of frequencies labelled in the inset. (**C**) Dots: Plots of the experimental SPP dispersion derived from fits to the line profiles show in (B) (see methods, Fig. S4) (*37*). The experimental data is superimposed on our theoretical model of Im($r_p$). Here, the absolute value of the graphene chemical potential |$\mu$| is used as a fitting parameter to the experimental graphene dispersion, yielding |$\mu$| = 0.61 eV (see methods, Fig. S5) (*37*).

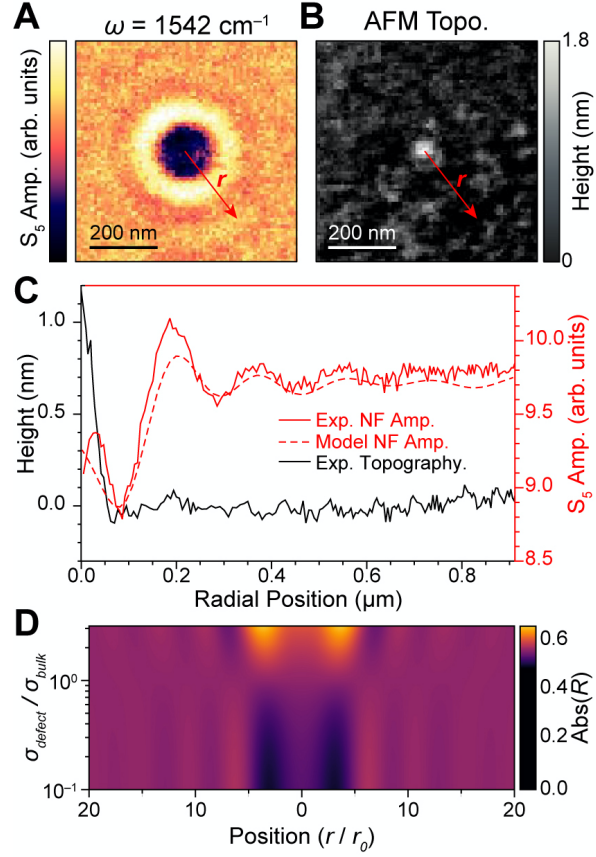

**Fig. 3. Analysis of SPP fringes near point defects in graphene/α-RuCl₃ heterostructures.** (**A**) Map of the near-field amplitude ($\omega$ = 1542 cm$^{-1}$) near a topographic point defect. (**B**) AFM-topography corresponding to the region in (A). (**C**) Solid red curve: Experimental near-field amplitude as a function of radial distance from the topographic point defect for $\omega$ = 920 cm$^{-1}$. Dashed red curve: the model fit to the experimental near-field profile based on the assumption of a Lorentzian charge-deficit with $r$ = 40 nm at x = 0. Solid black curve: radial line profile of AFM topography as a function of distance from the center of the defect. (**D**) Model radial fringe profile as a function of defect conductivity $\sigma_{defect}$ relative to the graphene bulk. Effective reflectance $R$ of illuminating fields from a dipole-like probe approximate the experimental near-field signal. The sign and magnitude of the first fringe are determined by magnitude of the defect conductivity relative to that of the bulk.

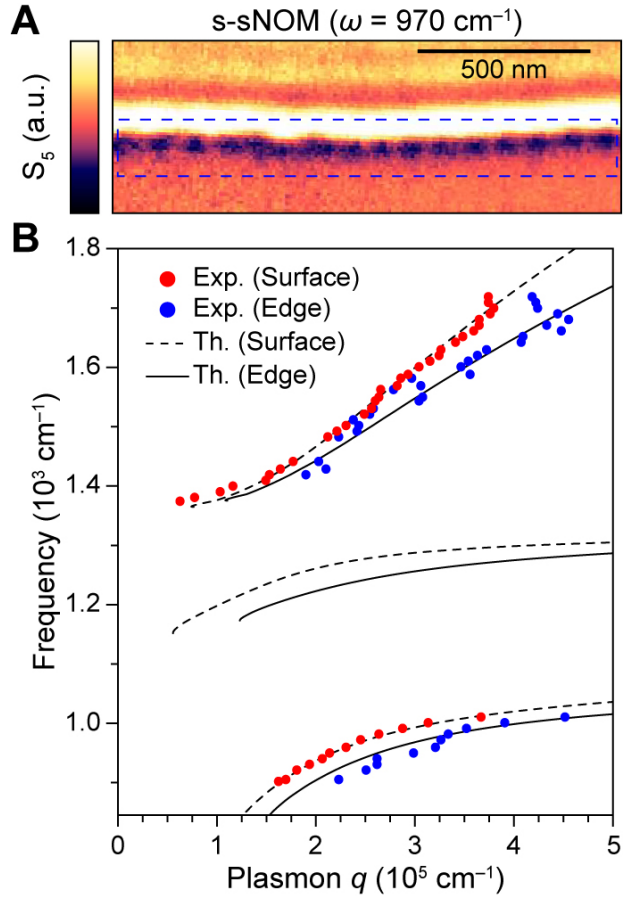

**Fig. 4. Analysis of EPP dispersion in graphene/α-RuCl₃ heterostructures.** (**A**) Characteristic s-SNOM image ($\omega = 970$ cm$^{-1}$) of EPP fringes along the graphene edge. (**B**) Red dots: The low-frequency SPP dispersion reproduced from Fig. 2B. Blue dots: The EPP dispersion extracted from lines profile of the near-field amplitude along the graphene edge (see methods, Fig. S4) (*37*). The dashed (solid) line shows the expected SPP (EPP) dispersion based on the assumption of a discontinuous jump in the graphene conductivity along that graphene edge (See supplementary materials) (*37*).

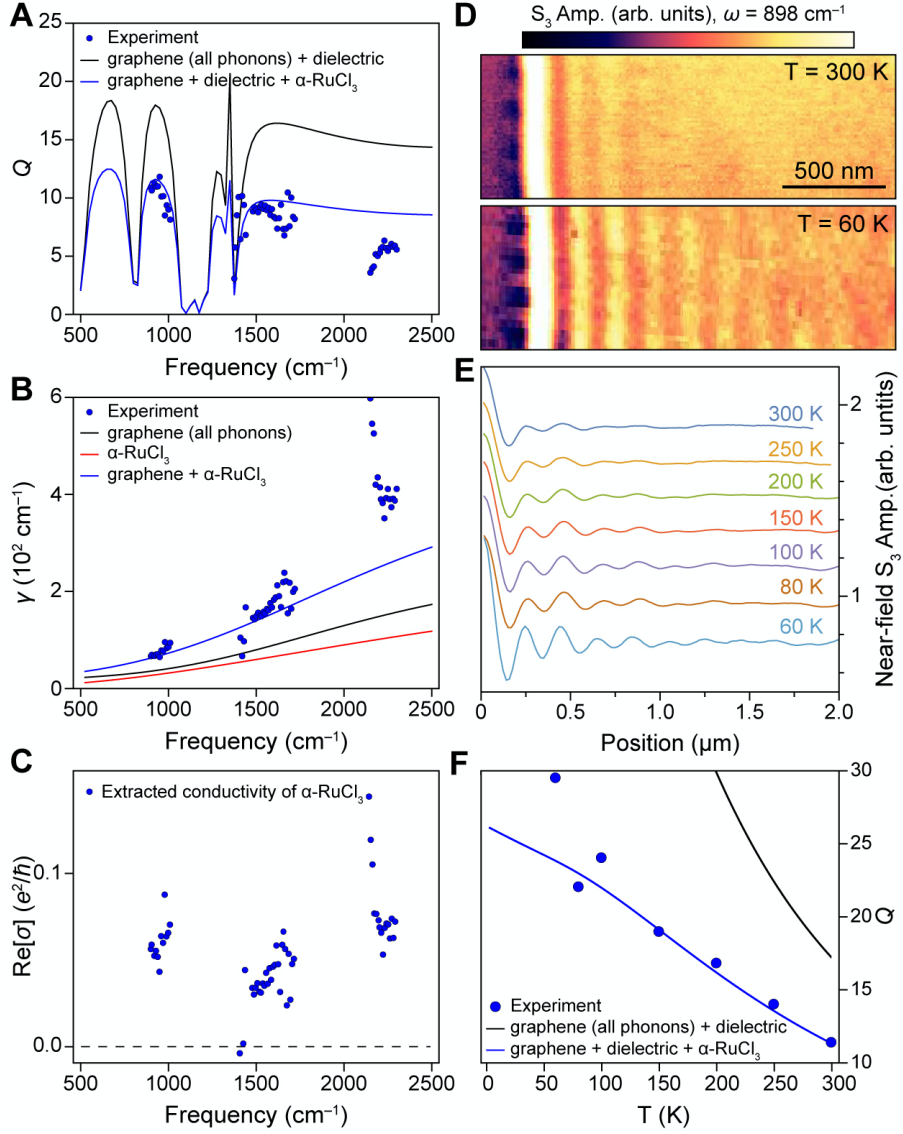

**Fig. 5. Temperature- and frequency-dependence of SPP losses in graphene/α-RuCl₃ heterostructures.** (**A**) Blue dots: Extracted quality factor, *Q*, for the edge-launched SPPs versus frequency. Black line: model frequency-dependent quality factor based on graphene phonon scattering and the dielectric environment only. Blue line: The model frequency-dependent quality factor based on graphene phonon scattering, the dielectric environment, and model losses to the interfacial α-RuCl₃ layer. (**B**) Blue dots: The extracted scattering rate versus frequency derived from the quality factor in (A) (*37*). Black line: The model frequency-dependent scattering rate

based on graphene phonon scattering only. Red line: The frequency-dependent contribution of the model α-RuCl$_3$ to the interfacial scattering. Blue line: The total frequency-dependent scattering of the graphene and α-RuCl$_3$ layer. (**C**) The extracted optical conductivity of interfacial α-RuCl$_3$ based on the excess scattering observed in (B) (*37*). (**D**) Map of the near-field amplitude ($\omega = 898$ cm$^{-1}$) near a graphene edge in graphene/α-RuCl$_3$ heterostructures taken at 300 K (top panel) and 60 K (bottom panel) under ultrahigh vacuum (UHV) conditions. (**E**) Line profiles of the average near-field amplitude as a function of distance from the graphene edge taken at the indicated sample temperatures ranging from 60 – 300 K. (**F**) Blue dots: The extracted SPP *Q* as a function of temperature taken from fits to the line profiles in (E) (see methods, Fig. S4) (*37*). Black line: the model temperature-dependent *Q* based on graphene phonons and dielectric losses only. Blue line: the model temperature-dependent *Q* based on graphene phonon scattering, the dielectric environment, and losses to the interfacial α-RuCl$_3$.